\documentclass[twocolumn, superscriptaddress, preprintnumbers, amsmath, amssymb, aps]{revtex4}

\allowdisplaybreaks
\allowdisplaybreaks[4]
\usepackage{graphicx}
\usepackage{dcolumn}
\usepackage{bm}
\usepackage{xurl}

\usepackage[pdfstartview=FitH,
            CJKbookmarks=true,
            bookmarksnumbered=true,
            bookmarksopen=true,
            colorlinks,
            linkcolor=blue,
            anchorcolor=blue,
            citecolor=blue
            ]{hyperref}

\usepackage{anyfontsize}
\usepackage{braket}
\begin{document}

\title{Collective quantum tunneling with time-dependent generator coordinate method}

\author{Wenmin Deng}
\affiliation{School of Physics and Astronomy, Beijing Normal University,
Beijing 100875, China}

\author{Guangping Chen}
\affiliation{School of Physics and Astronomy, Beijing Normal University,
Beijing 100875, China}

\author{Ganlong Ding}
\affiliation{School of Physics and Astronomy, Beijing Normal University,
Beijing 100875, China}

\author{Sibo Wang}
\affiliation{Department of Physics and Chongqing Key Laboratory for Strongly Coupled Physics, Chongqing University, Chongqing 401331, China}
\affiliation{Department of Physics, Graduate School of Science, The University of Tokyo, Tokyo 113-0033, Japan}

\author{Jing Peng}
\email[Corresponding author:~]{jpeng@bnu.edu.cn}
\affiliation{School of Physics and Astronomy, Beijing Normal University, Beijing 100875, China}
\affiliation{Key Laboratory of Multiscale Spin Physics (Ministry of Education), Beijing Normal University, Beijing 100875, China}

\author{Haozhao Liang}
\email[Corresponding author:~]{haozhao.liang@phys.s.u-tokyo.ac.jp}
\affiliation{Department of Physics, Graduate School of Science, The University of Tokyo, Tokyo 113-0033, Japan}
\affiliation{Quark Nuclear Science Institute, The University of Tokyo, Tokyo 113-0033, Japan}
\affiliation{RIKEN Center for Interdisciplinary Theoretical and Mathematical Sciences (iTHEMS), Wako 351-0198, Japan}

\date{\today}

\begin{abstract}
Inspired by the work of McGlynn and Simenel [Phys. Rev. C {\bf 102}, 064614 (2020)], this study investigates the quantum tunneling of two interacting distinguishable particles in two potential wells.
We first benchmark the system by reproducing key established results: the exact quantum solution and the spurious self-trapping effect that arises in the real-time mean-field dynamics for strong interactions.
To exactly capture the tunneling dynamics, we apply the time-dependent generator coordinate method (TDGCM) to the model.
Numerical simulations demonstrate that the TDGCM, by utilizing the real-time mean-field states as generator states, successfully overcomes the self-trapping effect, yielding tunneling dynamics in excellent agreement with the exact solution.
Furthermore, we explore the expectation values of the generator coordinates from the correlated TDGCM many-body wave function.
While different methods for calculating expectation values show consistent results in some cases, significant discrepancies are observed in others, providing critical insights into the emergence of collective and single-particle behaviors in interacting systems.
This work also verifies the TDGCM as a robust framework for describing collective quantum tunneling and opens avenues for its application to more complex and realistic systems.

\end{abstract}

\maketitle

\section{Introduction}

Quantum tunneling is a fundamental phenomenon in various fields of physics and chemistry~\cite{RevModPhys.62.251,RevModPhys.70.77} and represents one of the most important quantum mechanical effects~\cite{10.1063/1.466899}. Theoretical studies of the collective quantum tunneling in many-body problems, face formidable computational challenges due to the vast dimensionality of the Hilbert space~\cite{RevModPhys.71.1253,SIMENEL201819}.
This makes direct numerical solutions of the Schr\"odinger equation intractable, compelling the development of simplified theoretical models to investigate the underlying physical mechanisms.

In the development of computational many-body approaches for nuclear structure, two directions have been intensively explored: one is to pursue the exact solutions in relatively small model spaces, and the other is to search for mean-field approaches and their extensions in comparably large single-particle spaces~\cite{PhysRevC.103.064302}. The time-dependent Hartree-Fock (TDHF) method or its density functional theory (DFT) variant~\cite{RevModPhys.88.045004}, by replacing the many-body dynamical problem of interacting particles with the dynamics of independent particles moving in an average self-consistent mean field, is certainly among the most widely used tools to describe the properties of mesoscopic systems~\cite{Lacroix2014}.
However, mean-field theories describe systems using independent single-particle wave functions (e.g., a Slater determinant), neglecting the correlation effects between particles.
It yields a spontaneously symmetry-breaking ground state yet fails to restore the true symmetry of the system, while it may also overlook some crucial quantum fluctuations and correlations in specific phenomena and properties, including quantum tunneling~\cite{Lacroix2014,SIMENEL2012,RevModPhys.54.913,Bohr1975,HASEGAWA2020135693}. 

As clearly demonstrated in a recent work by McGlynn and Simenel~\cite{PhysRevC.102.064614}, the time-dependent Hartree method exhibits a self-trapping effect when dealing with strongly interacting systems: the system becomes confined by its own mean field, unable to penetrate the potential barrier, thus fundamentally failing to describe the quantum tunneling dynamics.
To overcome this problem, McGlynn and Simenel proposed an innovative imaginary-time mean-field approach.
By performing a Wick rotation that maps time onto the imaginary axis, they successfully enabled the mean-field evolution to explore the classically forbidden region in an exactly solvable two-well model.
The computed tunneling probabilities show an excellent agreement with exact solutions in the strong-interaction regime.
Their work opens up a new pathway for studying many-body tunneling phenomena.

Meanwhile, several efficient methods and algorithms for beyond-mean-field correlations have been developed in recent years.
These methods restore the symmetries broken by the static nuclear mean field and take into account the fluctuations around the mean-field minimum~\cite{PhysRevC.81.044311}.
The generator coordinate method (GCM)~\cite{Ring1980,P-GReinhard_1987,Schunck_2016} is one of the most effective microscopic approaches for treating collective motions and quantum fluctuations in quantum many-body systems, particularly in atomic nuclei.
In this framework, the system wave function is constructed as a superposition of the generator states, which are functions of generator coordinates, such as deformation parameters and the distance between two fragments.
In the domain of nuclear reactions, its time-dependent version, the time-dependent generator coordinate method (TDGCM) and its extensions have been developed and applied to predict the dynamics of heavy-ion collisions or fissions~\cite{10.3389/fphy.2020.00233,PhysRevC.105.044313,PhysRevC.108.014321,PhysRevC.111.L051302,PhysRevC.105.054604,PhysRevC.106.054609}.
Theoretically, TDGCM is also believed to overcome the self-trapping effect in quantum tunneling.
However, being an approximation itself, its final accuracy is directly influenced by factors such as the adequacy of correlation inclusion and the optimal selection of generator coordinates when solving the collective Schr\"odinger equation.
It is important to validate the reliability and precision of this method, particularly through a direct comparison with benchmark results.

Inspired by the work of McGlynn and Simenel~\cite{PhysRevC.102.064614}, the present study uses the identical two-well model with two interacting and distinguishable particles as a benchmark.
First, we revisit the exact solution of this model and the self-trapping effect arising from the real-time mean-field theory.
Second, we systematically construct the theoretical framework and computational scheme of the time-dependent generator coordinate method for this model.
We provide a detailed elaboration on the selection of collective coordinates, the construction of generator states, and the evolution of the collective wave function. 
Finally, we compare the numerical results obtained from TDGCM simulations with the exact solutions, in particular, to investigate the spurious self-trapping effect.
As a step further, we explore how the TDGCM wave functions can be utilized for describing the properties of the system within the conventional mean-field picture, for example, the expectation values of the generator coordinates.

\section{Theoretical framework}

\subsection{Two-well model} 

In line with the study by McGlynn and Simenel~\cite{PhysRevC.102.064614}, we consider the same two-well model containing two distinguishable particles that interact with each other. The two-well model offers a unique combination of analytical tractability and physical richness. Its exact solvability provides an unambiguous benchmark for assessing the accuracy of advanced many-body methods like the TDGCM. Despite its simplicity, the model incorporates the key ingredients of interaction-driven quantum dynamics, including the mean-field self-trapping effect and collective tunneling.

With two distinguishable particles labeled as $1$ and $2$, the system Hamiltonian is given by
\begin{equation}\label{Hamilt1}
    \hat{H}(1, 2) = \hat{h}_0(1) + \hat{h}_0(2) + \hat{V}(1, 2).
\end{equation}
A particle in the model can be described by two states $|L\rangle$ and $|R\rangle$, representing its position in the left and right wells, respectively.
The eigenstates of the single-particle Hamiltonian $\hat{h}_0$ are constructed as $| \pm \rangle = \frac{1}{\sqrt{2}} (|L\rangle \pm |R\rangle)$,  where \( |+\rangle \) denotes the symmetric ground state with energy 0 and $|-\rangle$ corresponds to the antisymmetric excited state. Then, the single-particle Hamiltonian is written as
\begin{equation}
    \hat{h}_0 = \alpha |-\rangle\langle-|, 
\end{equation}
where the potential barrier is increasing with decreasing $\alpha$. The interaction, which occurs only when both particles occupy the same well, is given by $\hat{V}(1, 2) = \mu (|LL\rangle \langle LL| + |RR\rangle \langle RR|)$, where $\mu$ controls the interaction strength, with positive (negative) values representing repulsion (attraction). The system is initialized in the $|LL\rangle$ state.

To establish a foundation for our subsequent work, we first revisit the model's exact solution and the spurious self-trapping effect that arises in the real-time mean-field dynamics, thereby reproducing the key results of McGlynn and Simenel.
For detailed derivations, readers are also referred to Ref.~\cite{PhysRevC.102.064614}. 

\subsection{Exact solution}

The exact evolution is determined from the time-evolution operator $\hat{U} = e^{-i\hat{H}t}$ (taking $\hbar = 1$). In the basis $\{|LL\rangle,|LR\rangle,|RL\rangle,|RR\rangle\}$, the Hamiltonian and the left‑well number operator read
\begin{equation}
\hat{H} = \begin{pmatrix}
    \alpha + \mu & -\alpha/2 & -\alpha/2 & 0 \\
    -\alpha/2 & \alpha & 0 & -\alpha/2 \\
    -\alpha/2 & 0 & \alpha & -\alpha/2 \\
    0 & -\alpha/2 & -\alpha/2 & \alpha + \mu
\end{pmatrix}
\label{Hamiltonian}
\end{equation}
and
\begin{equation}
\hat{N}_L = \begin{pmatrix} 
2 & 0 & 0 & 0 \\
0 & 1 & 0 & 0 \\
0 & 0 & 1 & 0 \\
0 & 0 & 0 & 0
\end{pmatrix},
\label{N_operator}
\end{equation}
respectively.
Starting from the initial state $|LL\rangle$, the state of the system at time $t$ is
\begin{equation}
    |\Psi(t)\rangle = \exp(-i\hat{H}t)|LL\rangle.\label{wave-function}
\end{equation}
Therefore, the expectation value of $\hat{N}_L$ in $|\Psi(t)\rangle$ can be expressed as
\begin{align}
    N_L(t) &= \langle \Psi(t)| \hat{N}_L |\Psi(t)\rangle \notag\\
    &= 1 + \frac{\beta - \mu}{2\beta} \cos\left( \frac{\beta + \mu}{2} t \right) + \frac{\beta + \mu}{2\beta} \cos\left( \frac{\beta - \mu}{2} t \right),\label{exact solution}
\end{align}
where $ \beta = \sqrt{4\alpha^2 + \mu^2}$. This expression explicitly shows the two oscillation modes that characterize the collective tunneling dynamics.

\subsection{Real-time mean-field}

For distinguishable particles, the time-dependent Hartree theory assumes the wave function remains in a product state, $|\Psi\rangle = |\psi_1\rangle \otimes |\psi_2\rangle$. Consequently, the two-body Hamiltonian is then approximated as $\hat{H}_H(1,2) = \hat{h}_H(1) + \hat{h}_H(2)$, with the time-dependent single-particle Hartree Hamiltonian defined by
\begin{equation}
\hat{h}_H(i) = \hat{h}_0(i) + \langle \psi_j | \hat{V}(1,2) | \psi_j \rangle,\qquad i \neq j.
\end{equation}
Since both particles are initially in the same state, they experience identical mean-field evolution. Therefore, they remain in the same single-particle state, $|\psi(t)\rangle = L(t)|L\rangle + R(t)|R\rangle$, which satisfies the time-dependent Hartree equation
\begin{equation}
    i \frac{\mathrm{d}}{\mathrm{d}t} |\psi(t)\rangle = \hat{h}_H(t) |\psi(t)\rangle.\label{Hatree}
\end{equation}

To represent the dynamics, the collective coordinates $\theta \in [-\pi/2, \pi/2]$ and $\phi \in [-\pi, \pi]$ are introduced via
\begin{equation}
    \theta = \arcsin (|L|^2 - |R|^2) \quad \text{and} \quad \phi = \arg (R/L). \label{define theta-phi}
\end{equation}
In these new coordinates, $L$ and $R$ are given by
\begin{equation}
L = \sqrt{\frac{1 + \sin \theta}{2}} e^{i \phi_L}, \quad R = \sqrt{\frac{1 - \sin \theta}{2}} e^{i \phi_R}, 
\end{equation}
where only the phase $\phi=\phi_R - \phi_L$ is physically relevant.
It is also seen that the normalization $|L|^2 + |R|^2 = 1$ is preserved. 

Inserting into Eq.~(\ref{Hatree}) and setting $\alpha = 1$ leads to a closed set of real-time mean-field equations,
\begin{subequations}\label{newmotion}
\begin{align}
    &\dot{\theta} = -\sin \phi, \label{newmotion1}\\
    \dot{\phi} = &\tan \theta \cos \phi + \mu \sin \theta.
    \label{newmotion2}
\end{align}
\end{subequations}
The average number of particles in the left well reads
\begin{equation}
    N_L(t)= 2 |L(t)|^2= 1 + \sin \theta(t).\label{real-time-expection}
\end{equation}

\subsection{Self-trapping} 

With the mean-field approximation and the choice of $\alpha=1$, the total (Hartree) energy is expressed as
\begin{align}
    \label{hartree_energy1}
    E &= K + U = \sum_{i=1}^{2} \langle \psi_i | \hat{h}_0 | \psi_i \rangle + \frac{1}{2} \sum_{i,j \neq i} \langle \psi_i \psi_j | \hat{V}(1,2) | \psi_i \psi_j \rangle\notag\\
    &=1 + \frac{\mu}{2} (1 + \sin^2 \theta) - \cos \theta \cos \phi,
\end{align} 
which is conserved under Eqs.~(\ref{newmotion1}) and~(\ref{newmotion2}), i.e., $d E/ d t =0$.

According to Eq.~(\ref{hartree_energy1}), the energy at  $\theta= \pm \pi / 2 $ is
\[
E_{1}\equiv\left.E\right|_{\theta= \pm \pi / 2}=1+\mu,
\]
while at  $\theta=0$  it is
\[
E_{2}(\phi)\equiv\left.E\right|_{\theta=0}=1+\frac{\mu}{2}-\cos \phi.
\]
A condition for the system to tunnel from one well to another in the real-time mean-field dynamics is the existence of $\phi$ satisfying $E_{2}(\phi)=E_{1}$; otherwise, the system is unable to cross the $ \theta=0 $ line. This equality requires $\mu / 2=-\cos \phi $, which is only possible when $|\mu| \leqslant 2$ . Self-trapping then occurs when  $|\mu| > 2$. This condition does not depend on the sign of  $\mu$. Thus, self-trapping occurs for both attractive and repulsive interactions at the same magnitude of the interaction strength.

\subsection{Time-dependent generator coordinate method}\label{TDGCM}

To accurately capture the tunneling between the two wells, we use TDGCM for this two-particle two-well system. More details for TDGCM can be referred to Refs.~\cite{Ring1980,10.3389/fphy.2020.00233}. 

Starting from a set of real-time evolution states that depend on $\theta$ with $\phi=0$, one assumes that the two-particle state of the system reads at any time
\begin{equation}
    |\Phi(t)\rangle = \sum_\theta |\Psi(\theta)\rangle g(\theta, t),
    \label{TDGCM wave function}
\end{equation}
where $g(\theta,t)$ are the weight functions and \(|\Psi(\theta)\rangle\) are the generator states taken as the product of two identical single-particle states from the real-time mean-field,
\begin{align}
    |\Psi(\theta)\rangle =& |\psi_1(\theta)\rangle \otimes |\psi_2(\theta)\rangle\notag\\
    =&\frac{1 + \sin \theta}{2} |LL\rangle+\frac{\cos \theta}{2} |LR\rangle\notag\\
    &+\frac{\cos \theta}{2}  |RL\rangle +\frac{1 - \sin \theta}{2}  |RR\rangle.
\end{align}
The overlap norm kernel and the Hamiltonian kernel are respectively defined as
\begin{align}
    \mathcal{N}(\theta, \theta^\prime) 
    &= \langle \Psi(\theta) | \Psi(\theta^\prime) \rangle\notag\\
    &=\frac{1+\sin\theta\sin\theta^\prime+\cos\theta\cos\theta^\prime}{2}
\end{align}
and
\begin{align}
    \mathcal{H}(\theta, \theta^\prime) =& \langle \Psi(\theta) | \hat{H} | \Psi(\theta^\prime) \rangle\notag\\
    =&(1+\mu)\frac{1+\sin\theta\sin\theta^\prime}{2}\notag\\
    &-\frac{\cos\theta+\cos\theta^\prime-\cos\theta\cos\theta^\prime}{2},
\end{align}
where we have set $\alpha = 1$ in accordance with the earlier convention.

The time evolution of $|\Phi(t)\rangle$ is governed by the Schr\"odinger equation 
\begin{equation}
\left( \hat{H} - i \frac{\mathrm{d}}{\mathrm{d}t} \right) |\Phi(t)\rangle = 0,
\end{equation}
which, when projected onto the generator states, leads to the time-dependent Griffin–Hill–Wheeler (GHW) equation
\begin{equation}
    \forall \theta^\prime: \quad \sum_\theta \left( \mathcal{H}(\theta^\prime, \theta) - i \mathcal{N}(\theta^\prime, \theta) \frac{\mathrm{d}}{\mathrm{d}t} \right) g(\theta, t) = 0.
\end{equation}

The initial condition corresponding to both particles in the left well, i.e., $|LL\rangle$ with $ (\theta,\phi) = (\frac{\pi}{2},0)$, translates into
\begin{equation}
    g(\theta, 0)=\delta_{\theta,\frac{\pi}{2}}.
\end{equation}
The expectation value of the left-well number operator $\hat{N}_L$ in the state $|\Phi(t)\rangle$ can then be evaluated as
\begin{align}
    N_L(t) =&\sum_{\theta,\theta^\prime} g^*(\theta,t)g(\theta^\prime,t)\langle \Psi(\theta) | \hat{N}_L |\Psi(\theta^\prime) \rangle\notag\\
    =&\sum_{\theta,\theta^\prime}\left[\frac{(1+\sin\theta)(1+\sin\theta^\prime)}{2}+\frac{\cos\theta\cos\theta^\prime}{2}\right]\notag\\
    &\times g^*(\theta,t)g(\theta^\prime,t).
\end{align}

\begin{widetext}
If the generator state is extended to include the phase variable  $\phi$ (with $\phi_L=0$), it takes a more general form as
\begin{align}
    |\Psi(\theta,\phi)\rangle &= |\psi_1(\theta,\phi)\rangle \otimes |\psi_2(\theta,\phi)\rangle\notag\\
    &= \frac{1 + \sin \theta}{2}  |LL\rangle+\frac{\cos \theta}{2} e^{i\phi} |LR\rangle +\frac{\cos \theta}{2} e^{i\phi} |RL\rangle +\frac{1 - \sin \theta}{2} e^{i 2\phi} |RR\rangle,
\end{align}
where the single-particle states now depend explicitly on both $\theta$ and $\phi$.
Similarly, one can obtain the expectation value of $\hat{N}_L$ as 
\begin{align}
    N_L(t)
    =\sum_{\theta,\phi;\theta^\prime,\phi^\prime}\left[\frac{(1+\sin\theta)(1+\sin\theta^\prime)}{2}+\frac{\cos\theta\cos\theta^\prime}{2}e^{i(\phi^\prime-\phi)}\right] g^*(\theta,\phi,t)g(\theta^\prime,\phi^\prime,t).
\end{align}
\end{widetext}

Since the set of generator states is generally linearly dependent, the matrix $\mathcal{N}$ may possess zero eigenvalues. For numerical stability,  
we perform eigenvalue decomposition on \(\mathcal{N}\), force all eigenvalues smaller than the threshold \(\epsilon = 10^{-10}\) to be fixed to \(\epsilon\), and then reconstruct the matrix using the original eigenvectors. This regularization process ensures the positive definiteness and invertibility of the matrix, and effectively avoids the numerical divergence problem caused by small eigenvalues while retaining the original information as much as possible.

\subsection{Tunneling rate}\label{Tunneling rate}

In the exact solution, the tunneling rate is given by twice the oscillation frequency between the left and right wells.
Equation~(\ref{exact solution}) shows that this oscillation consists of two modes, and only the lower frequency is associated with the tunneling process, yielding an exact tunneling rate
\begin{equation}
    \eta =\frac{\sqrt{4\alpha^2 + \mu^2} - |\mu|}{\pi}.
\end{equation}
Differentiating with respect to \(\mu\) (assuming \(\mu > 0\), so \(|\mu| = \mu\)), we get
\begin{equation}
    \frac{\mathrm{d}\eta}{\mathrm{d}\mu} = \frac{1}{\pi} \left( \frac{\mu}{\sqrt{4\alpha^2 + \mu^2}} - 1 \right).
\end{equation}
Since \(\frac{\mu}{\sqrt{4\alpha^2 + \mu^2}} < 1\) holds for all \(\mu > 0\), it follows that \(\frac{\mathrm{d}\eta}{\mathrm{d}\mu} < 0\), meaning that \(\eta\) decreases as \(\mu\) increases.

\section{Results and Discussion}

\subsection{TDGCM with generator coordinate $\theta$}

Starting from a set of real-time evolution states that depend only on $\theta$ with $\phi=0$, the accuracy of the TDGCM depends critically on the number of generator states included. When three or more distinct values of $\theta$ are used, the generator states span all possible configurations describing the positions that the two particles may occupy. Notably, apart from the requirement that one of $\theta$ must be taken as $\theta=\pi/2$ to satisfy the initial condition, the remaining distinct values of $\theta$ can be chosen randomly yet yield the identical results.

\begin{figure*}
\centerline{
    \includegraphics[width=0.8\textwidth]{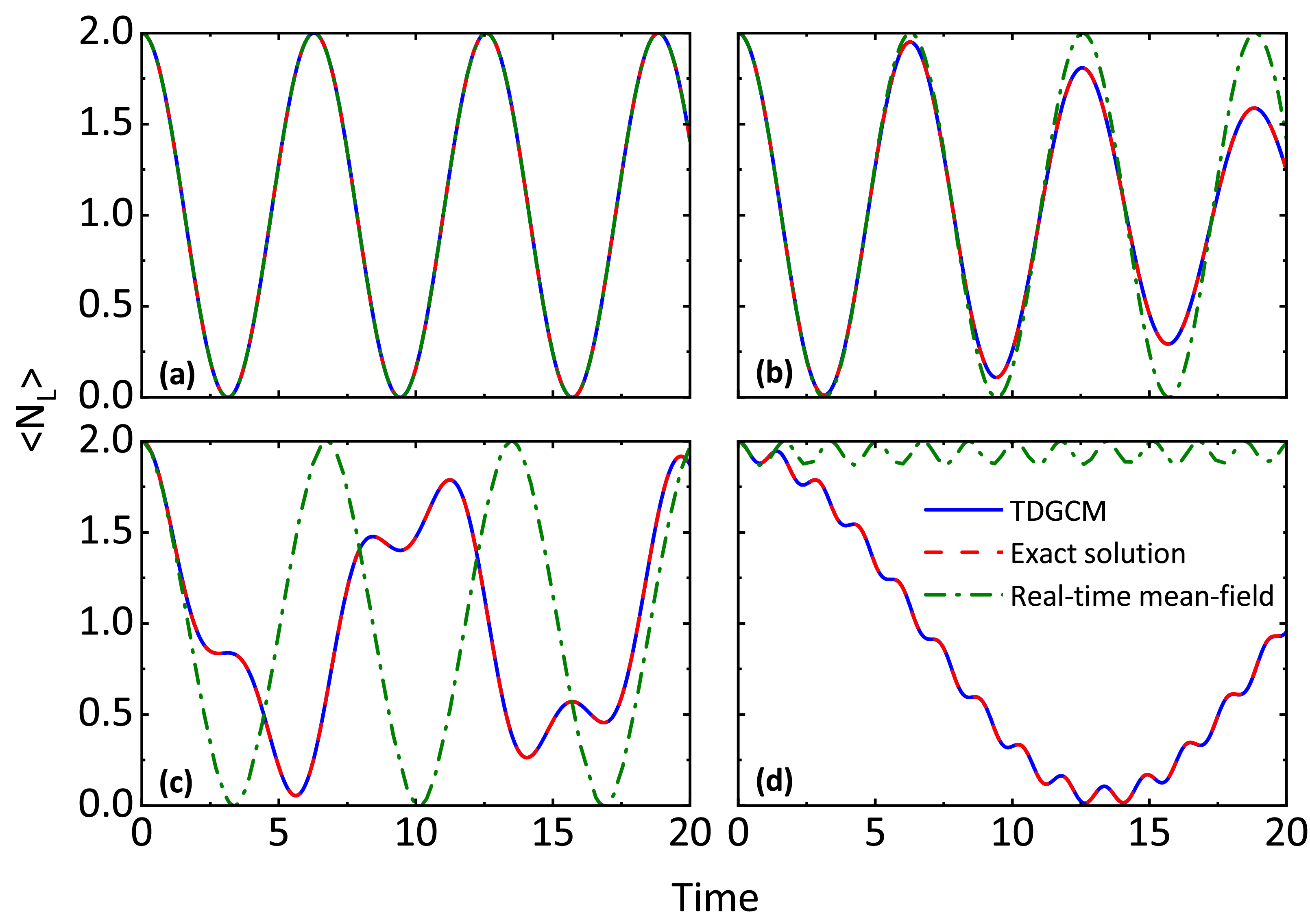}}
    \caption{TDGCM (blue solid lines), exact solutions (red dashed lines), and real-time mean-field (olive dash-dotted lines) predictions of $N_L(t)$ for (a) no interaction with $\mu=0$, (b) a very weak interaction with $\mu=0.1$, (c) a weak interaction with $\mu=1$, and (d) a strong interaction with $\mu=4$, where the number of generator states is $3$.} \label{three-generator-coordinate}
\end{figure*}

In Fig.~\ref{three-generator-coordinate}, the exact solutions and the TDGCM and real-time mean-field predictions of $N_L(t)$ are shown for the cases of no interaction with $\mu=0$, a very weak interaction with $\mu=0.1$, a weak interaction with $\mu=1$, and a strong interaction with $\mu=4$, where the number of generator states is $3$.
As demonstrated in the figure, the collective wave function accurately describes tunneling between the two wells. The results are in excellent agreement with the exact solutions across both weak- and strong-interaction regimes. The oscillatory behavior demonstrates the particles' ability to tunnel between the left and right wells. This tunneling process slows down as the interaction strength increases, as shown in Sec.~\ref{Tunneling rate}.
Moreover, when more than three distinct values of $\theta$ are used, the bases become overcomplete, but the results remain unchanged with an appropriate numerical treatment, discussed in Sec.~\ref{TDGCM}. 

\begin{figure}
  \begin{center}
    \includegraphics[width=0.95\linewidth]{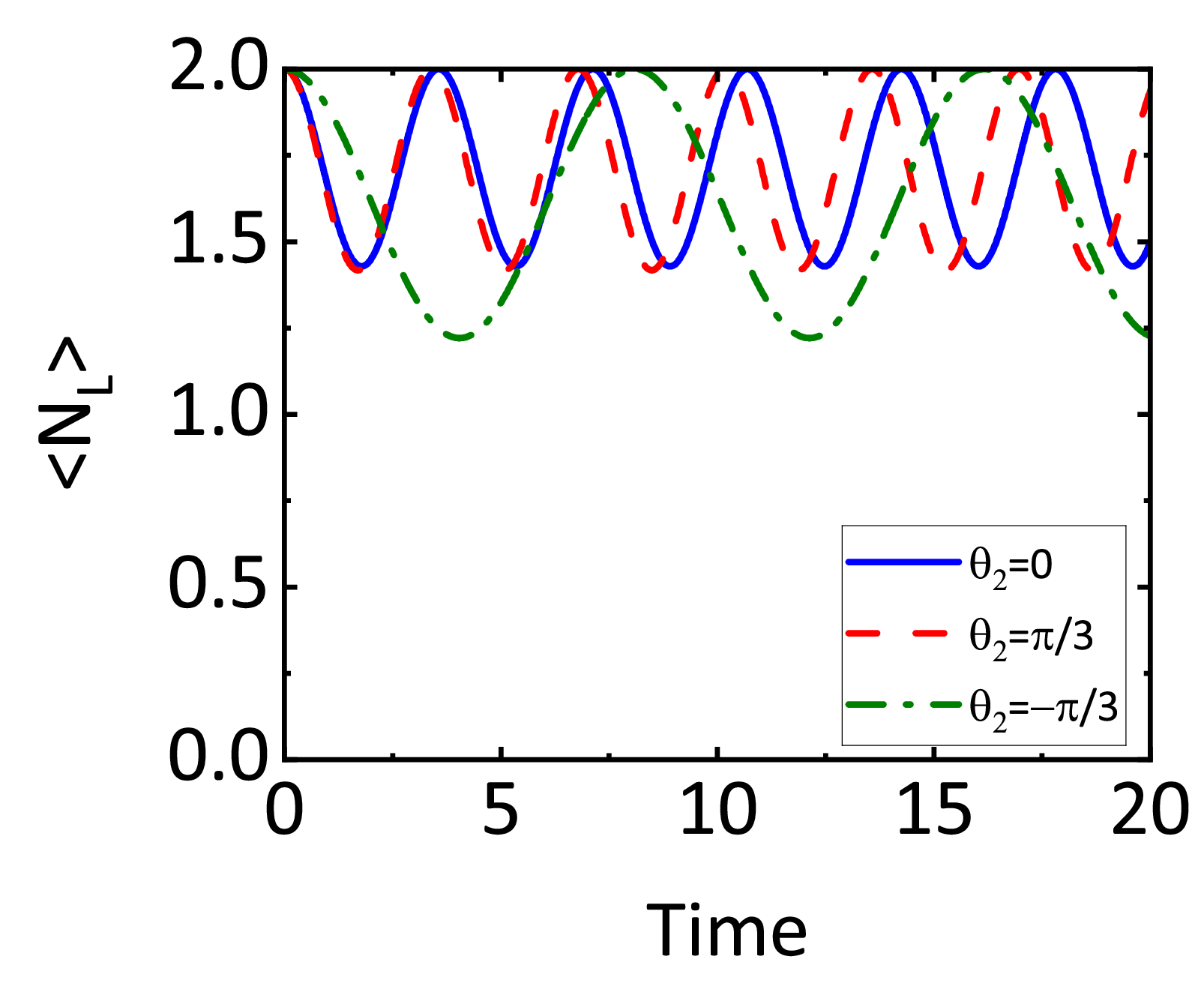}
    \caption{TDGCM predictions of $N_L(t)$ for the case of $\mu=1$.
    The results are calculated with only two generator states, and the value of $\theta_2$ is chosen as $\theta_2 = - \pi/3$, $0$, and $\pi/3$, respectively.}
\label{One-generator-coordinate}
  \end{center}
\end{figure}

In contrast,  the description fails if the number of bases is insufficient. For only two generator states, to satisfy the condition of the initial state, we must choose one generator coordinate as $\theta_1=\pi/2$ and the other one as an arbitrary $\theta_2$ to construct a set of generator states,
\begin{align}
    |\Psi(\frac{\pi}{2})\rangle =&|LL\rangle,\notag\\
    |\Psi(\theta_2)\rangle =&\frac{1 + \sin \theta_2}{2} |LL\rangle+\frac{\cos \theta_2}{2} (|LR\rangle+  |RL\rangle) \notag\\
    &+\frac{1 - \sin \theta_2}{2}  |RR\rangle.\label{two generator coordinate}
\end{align}
From Eq.~(\ref{two generator coordinate}), it is seen that the set of generator states cannot fully span the relevant configuration space. If $\theta_2=-\pi/2$, the generator states reduce to $|LL\rangle$ and $|RR\rangle$, which fail to describe tunneling. Even with other choices of $\theta_2$, the wave function fails to describe configurations such as simultaneous tunneling of both particles to the right well, as shown in Fig.~\ref{One-generator-coordinate} with the examples of $\theta_2 = - \pi/3$, $0$, and $\pi/3$. It is found that when the number of generator states is chosen as $2$, the results depend on the choice of $\theta_2$. The observation that results are independent of $\theta$ for three or more generator states, but depend on $\theta_2$ for two generator states, leads to the conclusion that if the results are independent of the choice of $\theta$, it indicates that the number of generator states is sufficient; conversely, if the results depend on the choice of $\theta$, it suggests that the basis is incomplete.

In a more extreme case, if one chooses only a single value of $\theta$, one can only choose $\theta=\pi/2$. The generator state collapses to $|LL\rangle$, and the wave function cannot describe any tunneling, effectively trapping the system in the left well.

The real-time mean-field dynamics show a critical dependence on the interaction strength $\mu$. While tunneling persists in weakly interacting regimes ($|\mu| < 2$), a transition occurs at the critical value $|\mu| = 2$. Beyond this threshold, the system exhibits a self-trapping phenomenon, where particles become confined to their initial well, and the tunneling is completely suppressed. This breakdown highlights the fundamental limitation of the real-time mean-field theory in describing strongly interacting quantum systems.

In contrast, TDGCM demonstrates superior robustness and accuracy.
In the non-interacting case ($\mu=0$), the TDGCM and real-time mean-field results coincide with the exact solution. 
As the interaction strength increases, the real-time mean-field results begin to deviate from the exact solution and the TDGCM predictions.
In the strong-interaction regime ($\mu>2$), where the real-time mean-field theory fails due to self-trapping, TDGCM continues to reproduce the exact tunneling dynamics faithfully, provided a sufficient number of generator coordinates is used. 

\subsection{TDGCM with generator coordinates $\theta$ and $\phi$}

Let us now consider the case that both $\theta$ and $\phi$ are taken as generator coordinates.
If both $\theta$ and $\phi$ are restricted to a single value, the initial condition forces the unique choice $\theta=\pi/2$ and $\phi=0$.
Consequently, the generator state collapses to $|LL\rangle$.

\begin{figure}
  \begin{center}
    \includegraphics[width=0.95\linewidth]{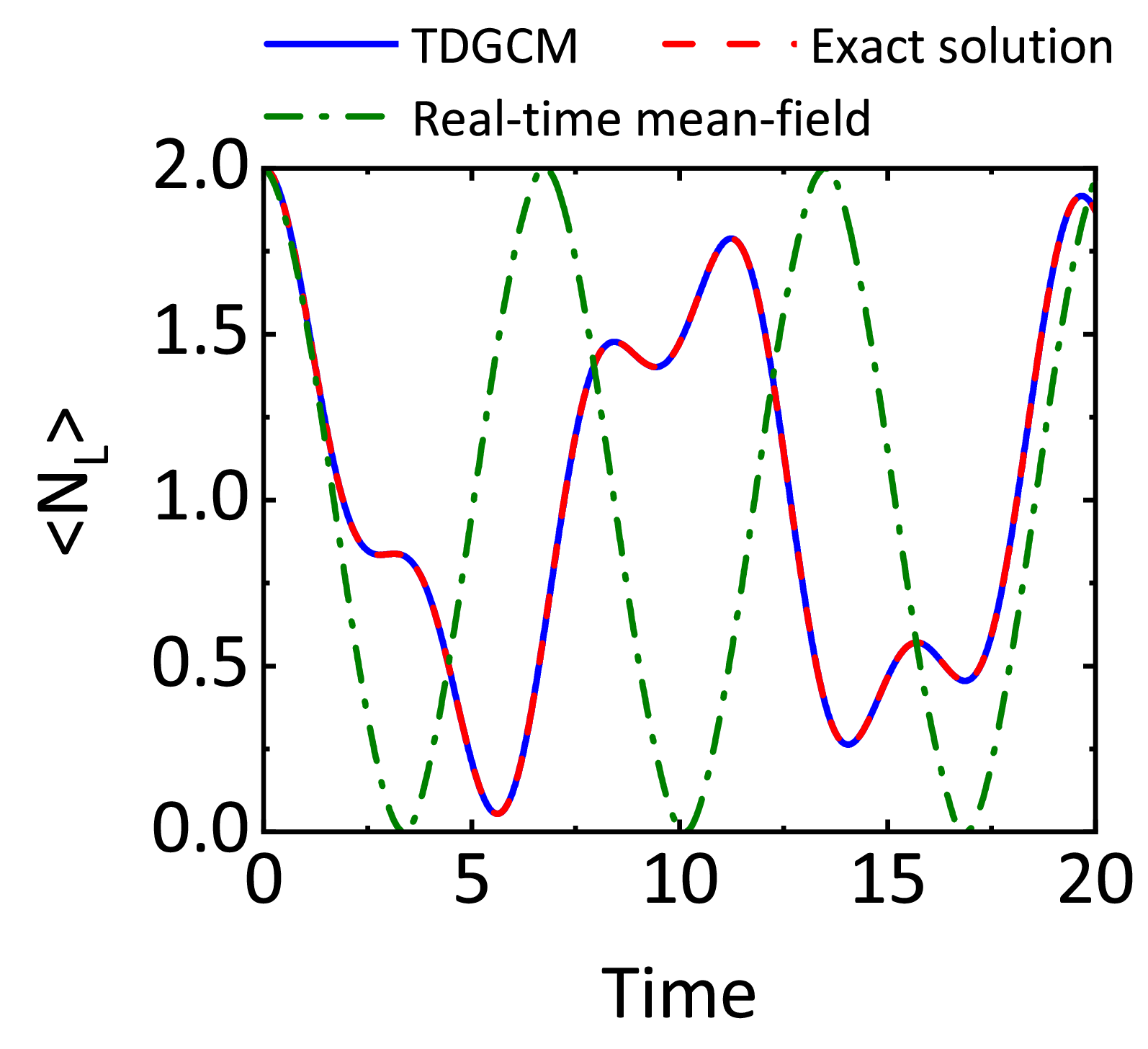}
    \caption{TDGCM (blue solid lines), exact solutions (red dashed lines), and real-time mean-field (olive dash-dotted lines) predictions of $N_L(t)$ for the case of $\mu=1$, where the numbers of both generator coordinates $\theta$ and $\phi$ are chosen as $2$.}
    \label{two-generator-coordinate}
  \end{center}
\end{figure}

If $\theta$ and $\phi$ are allowed to take two different values, for the same reason, one must choose $\theta_1=\pi/2$ and $\phi_1=0$ and the others as the arbitrary $\theta_2$ and $\phi_2$ to construct a set of generator states,
\begin{align}
    |\Psi(\frac{\pi}{2},0)\rangle =&|\Psi(\frac{\pi}{2},\phi_2)\rangle=|LL\rangle,\notag\\
    |\Psi(\theta_2,0)\rangle =&\frac{1 + \sin \theta_2}{2} |LL\rangle+\frac{\cos \theta_2}{2} (|LR\rangle+  |RL\rangle) \notag\\
    &+\frac{1 - \sin \theta_2}{2}  |RR\rangle,\notag\\
    |\Psi(\theta_2,\phi_2)\rangle =&\frac{1 + \sin \theta_2}{2}  |LL\rangle+\frac{\cos \theta_2}{2} e^{i\phi_2} (|LR\rangle+|RL\rangle)\notag\\
    &+\frac{1 - \sin \theta_2}{2} e^{i 2\phi_2} |RR\rangle.
    \label{two-two generator coordinate}
\end{align}
The generator states constructed in this way also include all possible configurations describing the positions that the two particles may occupy. However, such a set of states is generally linearly dependent, which causes the matrix $\mathcal{N}$ to possess zero eigenvalues. To ensure numerical stability, we constrain all eigenvalues to be greater than a small positive constant $\epsilon=10^{-10}$ as discussed in Sec.~\ref{TDGCM}. With this regularization, the collective wave function accurately captures tunneling between the two wells as shown in Fig.~\ref{two-generator-coordinate}.
We have also verified that $\theta_2$ and $\phi_2$ can be chosen randomly yet yield the identical results.

\section{Expectation values of generator coordinates}

On the one hand, in the theoretical description of many-body quantum tunneling and collective motion in the present model, the angular variables $\theta$ and $\phi$ serve as powerful collective coordinates that capture essential dynamical information of the system. Specifically, $\theta$ characterizes the population imbalance between two wells (e.g., $\sin\theta = |L|^2 - |R|^2$), while $\phi$ represents the relative phase between quantum states in the left and right wells ($\phi = \arg(R/L)$). The time evolution of $\langle\theta\rangle$ reveals whether the system can tunnel between configurations or becomes spuriously trapped due to mean-field approximations. When extended to imaginary time, $\theta$ and $\phi$ become complex quantities ($\theta = \theta_R + i\theta_I$, $\phi = \phi_R + i\phi_I$)~\cite{PhysRevC.102.064614}. Their expectation values along imaginary-time trajectories trace the classically forbidden path through the potential barrier, enabling quantitative calculation of tunneling probabilities.

On the other hand, the calculations of expectation values within the GCM have a rich history dating back to the foundational work of Hill, Wheeler, and Griffin~\cite{PhysRev.89.1102,PhysRev.108.311}. Reinhard and Goeke~\cite{P-GReinhard_1987,REINHARD1983141} demonstrated how the Gaussian overlap approximation (GOA) enables the mapping of microscopic operators to collective space, forming the foundation for calculating expectation values of observables such as multipole moments and transition probabilities. The calculation of expectation values has been extensively employed in the generator coordinate method (GCM) and its time-dependent extension (TDGCM) to extract crucial physical information about nuclear structure and dynamics. For instance, within the framework of the GCM, the expectation values of quadrupole deformation parameters $\beta$ and $\gamma$ have been systematically computed to analyze shape coexistence and triaxiality in magnesium isotopes, providing spectroscopic properties such as excitation energies and $B(E2)$ transition probabilities in excellent agreement with experimental data~\cite{PhysRevC.81.044311,PhysRevC.83.014308}. Furthermore, in fission studies, the expectation values of fragment observables have enabled a quantitatively microscopic description of fission dynamics within both the generalized and dissipative TDGCM frameworks~\cite{PhysRevC.105.044313,PhysRevC.108.014321,PhysRevC.111.L051302,PhysRevC.105.054604,PhysRevC.106.054609}. These successful applications motivate the present study, in which we emloy several methods to compute the expectation values of the angular variables $\theta$ and $\phi$ that characterize the population imbalance and relative phase in a two-well system, thereby providing a microscopic probe of many-body quantum tunneling dynamics.

Furthermore, to bridge the collective many-body dynamics with the underlying single-particle behaviors, we examine the generator coordinates of the system. Here, the generator coordinates $\theta$ and $\phi$, which are defined in Eq.~\eqref{define theta-phi} within the conventional mean-field picture, serve as ideal probes for this purpose. Calculating their expectation values directly from the many-body wave function, $\langle \theta \rangle$ and $\langle \phi \rangle$, allows us to investigate the single-particle characteristics from the collective many-body quantum state.

\subsection{The expectation values of $\theta$}

In general, many-body wave functions contain information beyond the mean-field picture.
For example, when we evaluate the expectation values of $\theta$ from the TDGCM wave functions in the present model, certain interpretations or model-dependent approximations are unavoidable.
Therefore, in the following, we employ several different definitions and computational schemes for cross-validation.

\subsubsection{Inversion method with mean-field approximation}

On the one hand, the number operator $\hat N_L$ in Eq.~\eqref{N_operator} and its expectation value $\langle \hat N_L\rangle$ are well-defined, regardless the calculations are performed by using the exact, TDGCM, or real-time mean-field wave functions.
On the other hand, using the real-time mean-field predictions of $N_L(t)$ given in Eq.~(\ref{real-time-expection}), we can obtain one of the possible connections between $\langle \theta\rangle$ and $\langle \hat N_L\rangle$, which reads
\begin{equation}
\langle \theta(t)\rangle=\arcsin(N_L(t)-1).
\end{equation}

The expectation value $N_L(t)$ calculated by the TDGCM wave functions reads
\begin{equation}
    N_L(t) =\langle \Phi(t) | \hat{N}_L |\Phi(t) \rangle = 2|c_{LL}(t)|^2+|c_{LR}(t)|^2+|c_{RL}(t)|^2,
\end{equation}
and thus we have
\begin{equation}
\langle \theta(t)\rangle=\arcsin(|c_{LL}(t)|^2-|c_{RR}(t)|^2),
\label{theta-inversion}
\end{equation}
where $c_{LL},c_{LR},c_{RL},c_{RR}$ are the coefficients in the expansion  $|\Phi(t)\rangle = c_{LL}(t)|LL\rangle + c_{LR}(t)|LR\rangle + c_{RL}(t)|RL\rangle + c_{RR}(t)|RR\rangle$, and similarly hereinafter.

\subsubsection{Density matrix method}

In the basis $\{|LL\rangle,|LR\rangle,|RL\rangle,|RR\rangle\}$, the operators that count whether particle 1 is in the left or right well read
\begin{equation}
    \hat{N}_{1L} = \begin{pmatrix} 
1 & 0 & 0 & 0 \\
0 & 1 & 0 & 0 \\
0 & 0 & 0 & 0 \\
0 & 0 & 0 & 0
\end{pmatrix},\quad 
\hat{N}_{1R} = \begin{pmatrix} 
0 & 0 & 0 & 0 \\
0 & 0 & 0 & 0 \\
0 & 0 & 1 & 0 \\
0 & 0 & 0 & 1
\end{pmatrix}.
\end{equation}
From the TDGCM state $|\Phi(t)\rangle$ shown in Eq.~(\ref{TDGCM wave function}), the matrix density is
\begin{equation}
    \rho(t)=|\Phi(t)\rangle\langle \Phi(t)|.\label{density-matrix}
\end{equation}
Then, one can calculate the expectation values of $ \hat{N}_{1L} $ and $ \hat{N}_{1R} $ as
\begin{equation}
    \langle \hat{N}_{1L} \rangle={\rm tr}(\rho\hat{N}_{1L})=|c_{LL}|^2 +|c_{LR}|^2
\end{equation}
and
\begin{equation}
    \langle \hat{N}_{1R} \rangle={\rm tr}(\rho\hat{N}_{1R})=|c_{RR}|^2 +|c_{RL}|^2,
\end{equation}
respectively.
Using the definition of the collective coordinate $\theta$, one can get
\begin{align}
    \theta &= \arcsin (|L|^2 - |R|^2) \notag\\
    &=\arcsin (\langle \hat{N}_{1L} \rangle - \langle \hat{N}_{1R} \rangle) \notag\\
    &=\arcsin(|c_{LL}|^2-|c_{RR}|^2).
\end{align}

\subsubsection{Reduced density matrix method}\label{Reduced density matrix method}

With the density matrix (\ref{density-matrix}), one can obtain the reduced density matrix for particle $1$ by taking the partial trace with respect to particle $2$, i.e.,
\begin{align}
    \rho(1)&=\langle L_2|\rho|L_2\rangle +\langle R_2|\rho|R_2\rangle \notag\\
    &= \begin{pmatrix} 
    |c_{LL}|^2 +|c_{LR}|^2 &  c_{LL}c^*_{RL}+c_{LR}c^*_{RR} \\
    c^*_{LL}c_{RL}+c^*_{LR}c_{RR} &  |c_{RR}|^2 +|c_{RL}|^2
    \end{pmatrix}.\notag
\end{align}
which leads to
\begin{align}
    \theta &= \arcsin (|L|^2 - |R|^2) \notag\\
    &=\arcsin (\rho(1)_{11}-\rho(1)_{22})\notag\\
    &=\arcsin(|c_{LL}|^2-|c_{RR}|^2).
\end{align}

It follows from the preceding analysis that for the expectation values of $\theta$, the three distinct computational approaches yield identical outcomes. This agreement serves as a strong validation of both the methods and the underlying physical picture. Given the consistency demonstrated among the methods, we present results from only one representative approach in the subsequent figures for clarity.

\subsubsection{Probability-based weighted average}

Since the set of $|\Psi(\theta)\rangle$ is, in general, linearly dependent, the norm kernel $\mathcal{N}$ possesses zero eigenvalues. To solve the GHW equation, it corresponds to a diagonalization of $\mathcal{N}$, i.e.,
\begin{equation}
\sum_{\theta'} \mathcal{N}(\theta, \theta') u_k(\theta') = n_k u_k(\theta), 
\label{diagonalization}
\end{equation}
where eigenvalues $n_k$ ($k = 1, 2, \ldots$) are positive semi-definite.

Retaining only the eigenstates with $n_k \not= 0$, there exists a set of normalized vectors in the Hilbert space,
\begin{equation}
|k\rangle = \frac{1}{\sqrt{n_k}} \sum_\theta u_k(\theta) |\Psi(\theta)\rangle , 
\end{equation}
which are orthogonal and called the natural states. They span the smallest Hilbert space, which contains all the generator states $|\Psi(\theta)\rangle$. 

The wave function can be expanded as
\begin{equation}
|\Phi\rangle = \sum_{k, \, n_k \neq 0} g_k |k\rangle, 
\end{equation}
with the corresponding weight function
\begin{equation}
g(\theta) = \sum_{k, \, n_k \neq 0} \frac{g_k}{\sqrt{n_k}} u_k(\theta). 
\end{equation}

The collective wave functions $p(\theta)$ are obtained from the norm kernel eigenvectors,
\begin{equation}
    p(\theta)=\sum_{k, \, n_k \neq 0} g_k u_k(\theta),
\end{equation}
which are orthonormal and
\begin{equation}
    \sum_\theta |p(\theta)|^2 =1.
\end{equation}
Therefore, $|p(\theta)|^2$ can be interpreted as a probability amplitude~\cite{PhysRevC.81.044311,PhysRevC.83.014308} and the expectation value of $\theta$ can be computed as
\begin{equation}
    \langle \theta \rangle= \sum_\theta \theta|p(\theta)|^2.
\end{equation}
We call this average the probability-based weighted average.

\subsubsection{Overlap-based weighted average}

The probability amplitude for finding the generator state $|\Psi(\theta)\rangle$ in the wave function $|\Phi\rangle$ is~\cite{Ring1980}  
\begin{align}
    \langle \Psi(\theta) | \Phi \rangle &=  \sum_{k, \, n_k \neq 0} g_k \frac{1}{\sqrt{n_k}} \sum_{\theta^\prime} u_k(\theta^\prime) \langle \Psi(\theta) |\Psi(\theta^\prime)\rangle\notag \\
    &= \sum_{k, \, n_k \neq 0} \sqrt{n_k}g_k u_k(\theta),
\end{align}
where Eq.~\eqref{diagonalization} has been employed.

This leads to an alternative estimator 
\begin{equation}
    \langle \theta \rangle= \sum_\theta \theta|\langle \Psi(\theta) | \Phi \rangle|^2.
\end{equation}
We call this average the overlap-based weighted average.

\subsubsection{Eigenvalue-based weighted average}

If $\theta$ is regarded as the eigenvalue of operator $\hat{\theta}$ on $|\Psi(\theta)\rangle$, one can calculate the expectation value of $\hat{\theta}$ as
\begin{align}
    \langle \hat{\theta} \rangle&= \langle\Phi|\theta|\Phi\rangle\notag\\
    &=\sum_{\theta,\theta^\prime} \theta g^*(\theta^\prime, t)g(\theta, t)\langle\Psi(\theta^\prime)|\Psi(\theta)\rangle\notag\\
    &=\sum_{\theta,\theta^\prime} \theta g^*(\theta^\prime, t)g(\theta, t)\mathcal{N}(\theta^\prime,\theta).
\end{align}
We call this average the eigenvalue-based weighted average.

\begin{figure*}
  \begin{center}
    \includegraphics[width=0.8\textwidth]{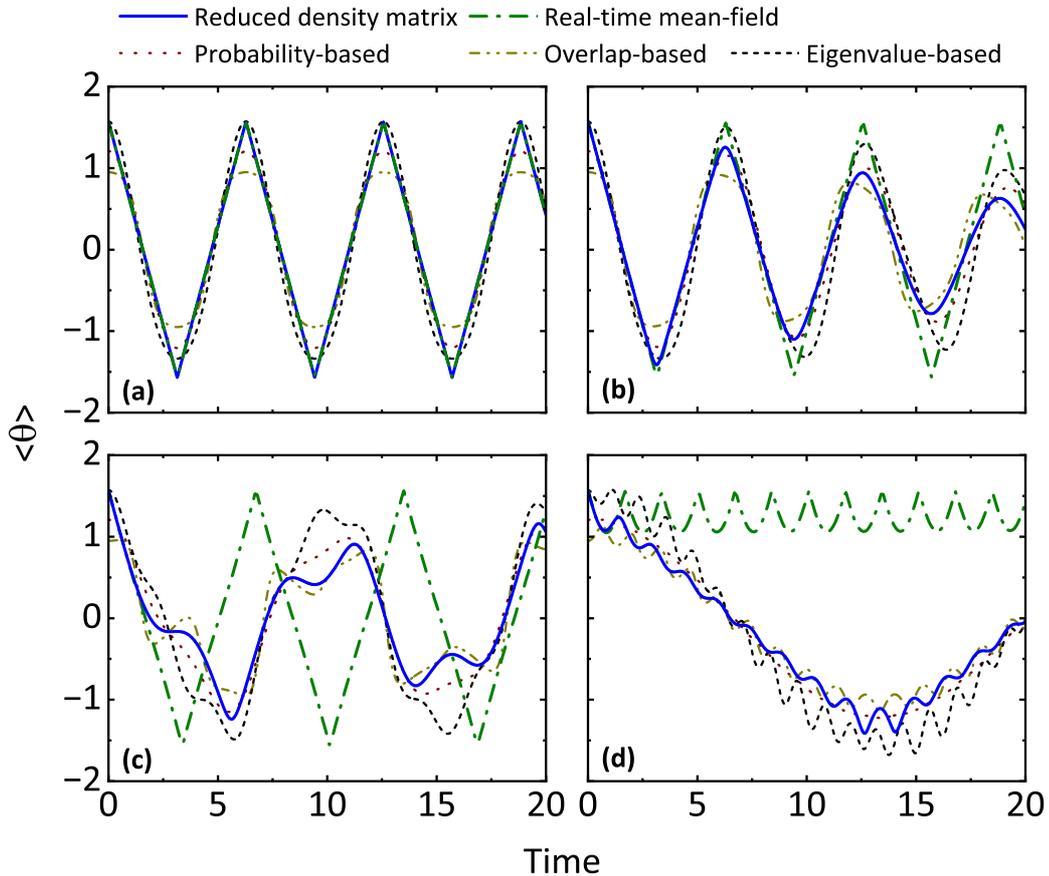}
    \caption{Evaluations of $\langle\theta(t)\rangle$ calculated by the reduced density matrix (blue solid lines), real-time mean-field (olive dash-dotted lines), probability-based weighted average (cyan dotted lines), overlap-based weighted average (yellow dark dash-dot-dotted lines), and eigenvalue-based weighted average (black short-dashed lines) for (a) no interaction with $\mu=0$, (b) a very weak interaction with $\mu=0.1$, (c) a weak interaction with $\mu=1$, and (d) a strong interaction with $\mu=4$.}\label{theta}
  \end{center}
\end{figure*}

\subsubsection{Real-time mean-field approach}\label{phi real-time}

Within the real-time mean-field approach, the quantities $\theta$ and $\phi$ are defined and satisfy a closed set of equations~\eqref{newmotion}.
Numerical integration of these equations yields $\theta(t)$ directly.
Note that the real-time mean-field approach becomes exact and serves as a benchmark for the case without interaction, i.e., $\mu=0$.

\subsubsection{Results and discussion}

In Fig.~\ref{theta}, the evaluations of $\langle\theta(t)\rangle$ calculated by the reduced density matrix, real-time mean-field, probability-based weighted average, overlap-based weighted average, and eigenvalue-based weighted average are shown for the cases of no interaction with $\mu=0$, a very weak interaction with $\mu=0.1$, a weak interaction with $\mu=1$, and a strong interaction with $\mu=4$.

As clearly shown in Fig.~\ref{theta}, the evaluations by several different methods all indicate the oscillatory behavior, which confirms that the particles can tunnel between the left and right wells. This tunneling process slows down as the interaction strength increases, reflected in the delayed arrival of particles in the right well, which is consistently observed in Fig.~\ref{three-generator-coordinate}. The agreement among different evaluations suggests these methods are able to capture the same underlying physical behavior through distinct mathematical formalisms.

Despite the success of capturing the same underlying physical behavior through different methods, the present comprehensive analysis reveals a dichotomy in the results. The inversion method with mean-field approximation, the density matrix method, and the reduced density matrix approach demonstrate remarkable consistency in their predictions, as already shown in theoretical analysis. 
Notably, in the non-interacting case ($\mu=0$), these three methods are in perfect agreement with the results of the direct real-time mean-field integration, which serves as a benchmark. However, as the interaction strength $\mu$ increases, the predictions from the direct mean-field integration begin to deviate from the other three methods, with the discrepancy growing systematically. In the strong-interaction regime ($\mu>2$), the direct real-time mean-field integration fails qualitatively due to the self-trapping effect discussed above, a breakdown not observed in the other approaches. 

In contrast, the three different kinds of weighted averages produce different results.
In particular, for the non-interacting case with $\mu=0$, the overlap-based weighted average exhibits the largest deviations from the exact expectation values, while the probability-based average shows smaller, yet significant, discrepancies.
The eigenvalue-based weighted average, though the most stable among the three, still demonstrates non-negligible deviations. All three weighted schemes display pronounced sensitivity to the variation of the interaction strength parameter $\mu$. 

This methodological divergence highlights several critical insights. First, the agreement among the first three methods validates their reliability for studying the system's properties. Second, the difference of the weighted average methods to their respective implementation details underscores how specific methodological choices can introduce significant biases in the interpretation of many-body effects.
The systematic nature of these discrepancies suggests they originate from fundamentally different treatments of the quantum system rather than numerical errors.

These findings carry important implications for methodological selection in many-body analysis. While the consistent methods provide a robust benchmark for basic properties, the varied performance of different weighted average schemes indicates they may encode different physical assumptions about the system's structure. This comprehensive comparison provides valuable guidance for future analysis of quantum many-body systems within the conventional mean-field picture.

\subsection{The expectation values of $\phi$}

It is tricky to evaluate the expectation values of $\phi$.
For example, if we take only $\theta$ as the generator coordinate, with $\phi=0$, all weighted-average methods yield $\langle \phi \rangle=0$. However, this result is incorrect, as will be clarified by the alternative methods described below.

\subsubsection{Inversion method with mean-field approximation}

On the one hand, the real-time mean-field method provides a first-order differential equation~\eqref{newmotion1} to relate the collective variables $\theta$ and $\phi$.
On the other hand, using the inversion method with mean-field approximation for $\langle \theta(t)\rangle$ in Eq.~\eqref{theta-inversion}, we can obtain an expression for $\langle \phi \rangle$ as
\begin{align}
    \langle\phi\rangle &=-\arcsin(\dot{\langle\theta\rangle})\notag\\
    &=-\arcsin\left(\frac{2\mathrm{Re}\big(c_{LL}^*\dot{c}_{LL} - c_{RR}^*\dot{c}_{RR}\big)}{\sqrt{1 - (|c_{LL}|^2-|c_{RR}|^2)^2}}\right).
\end{align}

\subsubsection{Reduced density matrix method}

Combining the reduced density matrix in Sec.~\ref{Reduced density matrix method} with the definition of $\phi$ in Eq.~\eqref{define theta-phi}, one can derive
\begin{align}
    \langle\phi\rangle &= \arg (R/L) =\arg (\rho(1)_{22}/\rho(1)_{12})\notag\\
    &=\arg\left(\frac{(|c_{RR}|^2+|c_{RL}|^2)}{c_{LL}c^*_{RL}+c_{LR}c^*_{RR}}\right).
\end{align}

Remarkably, we now obtain different results of $\langle\phi\rangle$ from the two methods described above.
This is different from the discussions of $\langle\theta\rangle$, where the results are consistent among the inversion method, density matrix method, and reduced density matrix method.

\subsubsection{Real-time mean-field approach}

As already shown in Sec.~\ref{phi real-time}, $\phi(t)$ can be obtained by numerically solving a closed set of equations~\eqref{newmotion} in the real-time mean-field approach.
This again serves as a benchmark for the non-interacting case.

\subsubsection{Results and discussion}

\begin{figure}
  \begin{center}
    \includegraphics[width=0.95\linewidth]{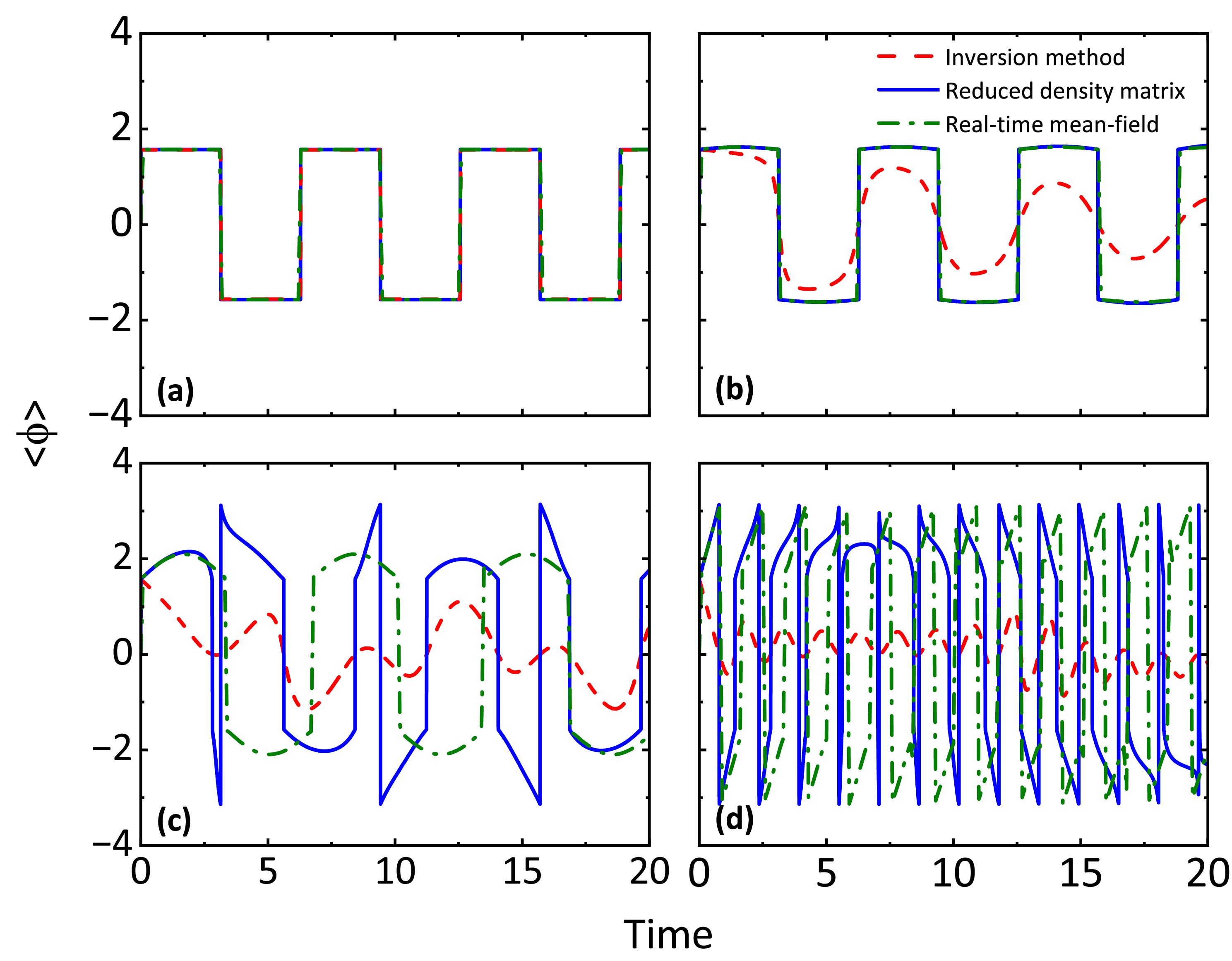}
    \caption{Evaluations of $\langle\phi(t)\rangle$ calculated by the inversion method with mean-field approximation (red dashed lines), reduced density matrix method (blue solid lines), and real-time mean-field (olive dash-dotted lines) for (a) no interaction with $\mu=0$, (b) a very weak interaction with $\mu=0.1$, (c) a weak interaction with $\mu=1$, and (d) a strong interaction with $\mu=4$.}\label{phi}
  \end{center}
\end{figure}

In Fig.~\ref{phi}, the evaluations of $\langle\phi(t)\rangle$ calculated by the inversion method, reduced density matrix, and real-time mean-field are shown for the cases of no interaction with $\mu=0$, a very weak interaction with $\mu=0.1$, a weak interaction with $\mu=1$, and a strong interaction with $\mu=4$.
As shown in the figure, the three methods exhibit a clear discrepancy in their results, indicating a systematic deviation that warrants further investigation.
While all methods are in agreement for the non-interacting case ($\mu=0$), their predictions gradually deviate from each other as the interaction strength $\mu$ increases. 
Notably, at $\mu=0.1$, the results obtained from the direct real-time mean-field integration remain consistent with those of the reduced density matrix method, in marked contrast to those from the inversion method. For $\mu=1$, the predictions of real-time mean-field and the reduced density matrix method exhibit distinct oscillation periods, while their amplitudes are comparable. In contrast, the inversion method yields a smooth curve, whereas the other two methods often exhibit boundary-related irregularities in their trajectories. With a strong interaction ($\mu=4$), the results obtained from both direct real-time mean-field integration and the reduced density matrix method oscillate rapidly with similar amplitudes, whereas the amplitude of those from the inversion method is significantly smaller.

These observations suggest that extracting single-particle phase information from a collective many-body wave function remains a nontrivial challenge. Among the methods compared, the reduced density matrix approach appears to offer a more robust prediction of the phase, likely owing to its foundation in the system's exact many-body wave function. Future work should aim to clarify the origin of these methodological differences and to develop more robust frameworks for resolving fine-grained, single-particle properties within collective tunneling dynamics.

\section{CONCLUSIONS}

In this study, we investigate the dynamics of collective quantum tunneling using a solvable two-well model with two interacting particles. We first revisit the exact solution of the model and reproduce a previously reported finding: namely, that the real-time mean-field dynamics exhibit an unphysical self-trapping effect beyond a critical interaction strength. We then systematically develop both the theoretical framework and the computational implementation of the time-dependent generator coordinate method (TDGCM) for this model, providing a detailed treatment of the selection of generator coordinates, the construction of generator states, and the evolution of the collective wave function.

In the present TDGCM calculations, the generator states are constructed from real-time mean-field solutions, and the resulting dynamics show remarkable agreement with the exact solutions, thereby successfully overcoming the self-trapping limitation. Furthermore, when extracting single-particle properties from the many-body wave function, we observe the same underlying physical behavior through different mathematical formulations, while also identifying a methodological dichotomy in the numerical results: some computational approaches consistently reproduce the benchmark values, whereas others show noticeable deviations. This discrepancy provides valuable insight into how different techniques describe the emergence of single-particle behavior from an underlying many-body state.

This work opens several promising directions for future research. A natural extension is to apply the same TDGCM framework to systems with more particles or with different types of interactions. More broadly, the present analysis contributes to a deeper understanding of the interplay between collective and single-particle degrees of freedom in correlated quantum systems.

\section*{ACKNOWLEDGMENTS}
This research was supported by the Super Computing Center of Beijing Normal University. 
This work was also supported by China Scholarship Council (CSC) (File No. [202406050068]) and National Natural Science Foundation of China (NSFC) under Grant No.~12575130. 

\bibliography{reference}

\end{document}